# Delayed 1T to 2H Phase Transition Upon Electrochemical Delithiation of LiMoS$_2$

Yerin Hong[1], Juhwan Lim[1], Jinhong Min[1], Nishkarsh Agarwal[1], Robert Hovden[1,3], Ageeth A. Bol[1,2,3], Yiyang Li[1†]

[1]Materials Science and Engineering, University of Michigan, Ann Arbor, MI 48109
[2]Chemistry, University of Michigan, Ann Arbor, MI 48109
[3]Applied Physics Program, University of Michigan, Ann Arbor, MI 48109

**ABSTRACT**.

Molybdenum disulfide (MoS$_2$) is a widely studied layered material for electronic, optical, and catalytic applications. It can host lithium ions between the van der Waals layers, which triggers a phase transition between the semiconducting 2H phase and metallic 1T phase. While lithium insertion triggers a phase transition to the 1T phase, the phase behavior upon electrochemical lithium removal is not resolved. In this work, we conduct single-flake electrochemical (de)lithiation of MoS$_2$ using microelectrode arrays. Through both electrochemical voltage analysis and correlative Raman spectroscopy, we show that an electrochemically cycled and delithiated MoS$_2$ flake initially remains in the 1T phase. However, over the course of several days, it transitions back into the thermodynamically stable 2H phase. This result resolves the phase transformation pathway upon delithiation and showcases the ability to electrochemically synthesize the metastable 1T-MoS$_2$ phase.

## I. INTRODUCTION.

MoS$_2$ is a well-studied transition metal dichalcogenide (TMD) comprising of two-dimensional planar atomic sheets held together by van der Waals forces. Applications of MoS$_2$ include lubricants [1], transistors [2–4], memory [5–7], batteries [8–10], catalysts [11–13], and water purification [14].

One important property of MoS$_2$ is its ability to host alkali metal ions like Li$^+$ within its van der Waals gaps. This triggers several changes in its physical properties which can be harnessed for diverse applications [15]. First, the chemical bonds formed between the MoS$_2$ and the alkali metal ions can be used for electrochemical energy storage. This enables its usage as a battery electrode, including one of the first Li-ion batteries by Moli Energy in the 1980s [16]. Second, the electrochemical insertion and removal of alkali metal ions change the van der Waals gap distance, which could be utilized in electrochemical actuators [17,18] and thermal transistors [3,19]. Finally, lithium ion insertion triggers phase transformation from the 2H semiconductor phase with a trigonal prismatic Mo-S coordination to a 1T metallic phase (including both 1T and distorted 1T$'$ phase) with an octahedral Mo-S coordination; this 1T/1T$'$ phase has the potential to form Ohmic contacts to other 2D semiconductors [2], catalysis [11], and has applications in neuromorphic computing [7,20,21].

Two types of lithium-ion intercalation into MoS$_2$ have been utilized: chemical and electrochemical. Chemical intercalation introduces lithium into the van der Waals layers through evaporation or through exposure to the liquid organometallic reagent n-butyllithium [22–26]. These synthetic methods are simple and scalable but they lack precise control over the amount of lithium ion inserted [15,27].

Alternatively, electrochemical insertion provides a more controlled, dynamic, and reversible approach to control the lithium concentration [28]. During electrochemical (de)intercalation, Li ions are inserted into and removed from MoS$_2$ through a liquid electrolyte. Due to charge balance, the applied electrochemical current equals the number of Li ions transferred. This not only enables precise control over the number of Li ions inserted but also enables a gentle process to remove Li ions and reverse the reaction [29]. Due to this precision and reversibility, electrochemical intercalation is needed for functional devices like batteries, actuators, and thermal transistors that require reversible modulation of the lithium concentration or the phase.

Understanding the dynamics of the reversible 2H to 1T/1T$'$ phase transformation is critical for engineering electrochemical devices. During electrochemical lithium intercalation, it is well known that the 2H-MoS$_2$ transforms into the 1T/1T$'$ phase, with a characteristic voltage plateau around 1.1 V vs Li/Li$^+$, indicating phase transformation [27,30–33]. However, the reverse dynamics are unclear. Some works suggest that the phase transition process is reversible, and that the 1T/1T$'$ phase transforms back into the thermodynamically stable 2H phase upon lithium removal [15,34,35]. Other papers utilizing porous electrodes suggest that MoS$_2$ is trapped in the metastable 1T/1T$'$ state even after Li removal [31,36–38]. This metastable 1T/1T` phase is also observed after deintercalating Li from Li$_x$MoS$_2$ (0<X<1) by oxidizing with water [38,39]. In addition, there are papers that report on MoS$_2$ decomposition during lithiation, and the reverse reaction results from Li-S redox [9,35,40] . Due to these conflicting results, the phase transformation dynamics during electrochemical delithiation are not clear. One challenge is that most electrochemical measurements have been conducted on porous electrodes containing many $MoS_2$ particles [35,41], making it difficult to elucidate the intrinsic phase transformation dynamics.

†Contact author: yiyangli@umich.edu

In this work, we conduct single-flake, current-controlled electrochemical intercalation of $MoS_2$ to investigate phase transformation dynamics during lithiation and delithiation. Our results show that 2H-$MoS_2$ transforms into 1T/1T′ during the first lithiation cycle, as previously reported [27,30–33]. However, we show that the flake remains in the metastable 1T/1T′ state immediately after delithiation using detailed electrochemical voltage trace analysis and correlative Raman spectroscopy. Then, over the course of several days, the metastable 1T/1T′ -$MoS_2$ slowly transforms back into the stable 2H phase. We provide the first demonstration of the ability to electrochemically create a single 1T/1T′ delithiated $MoS_2$ flake, as well as definitive evidence of its reversion back into the 2H phase over time in the absence of oxidizing chemicals. For the remainder of the main text, we will use 1T to represent both the 1T and distorted 1T′ phases as they could not be distinguished using the methods of this work.

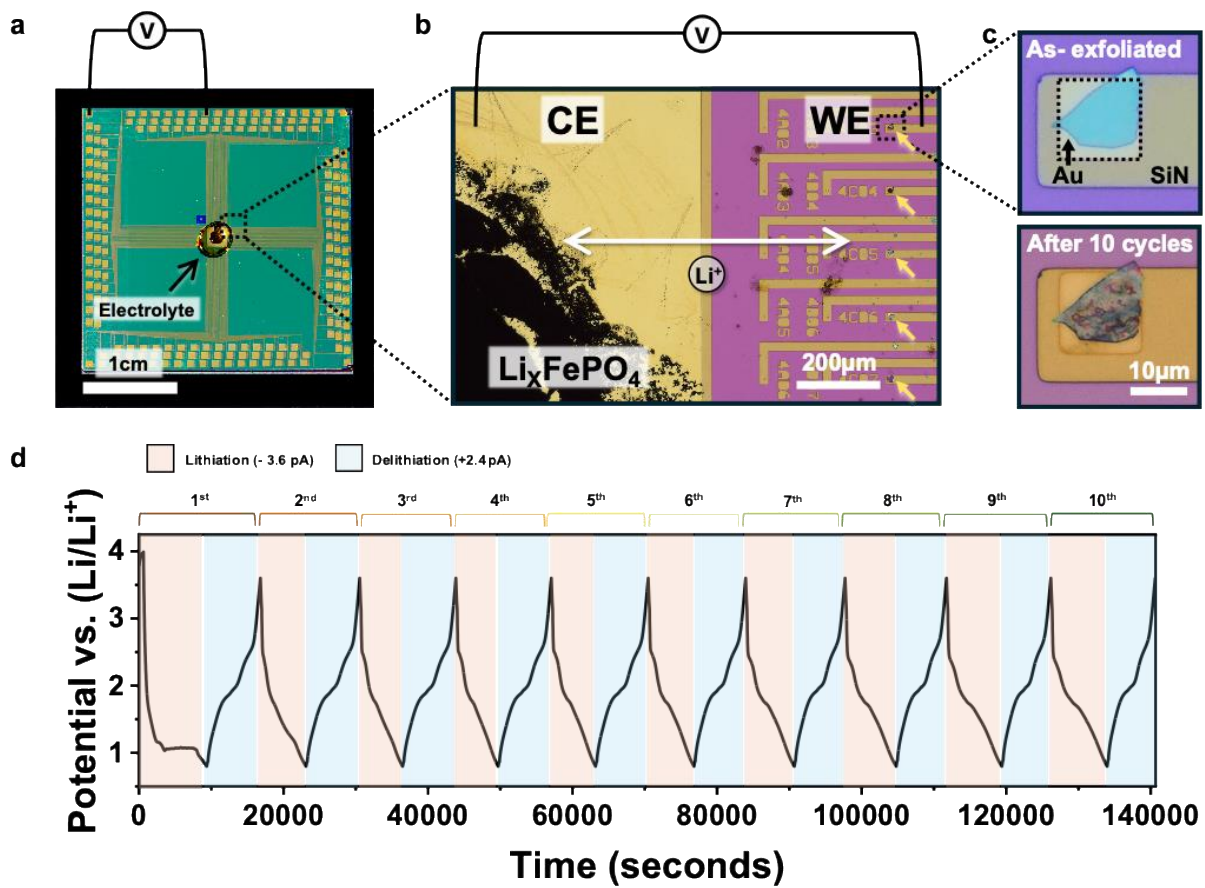

**Figure 1. Microelectrode array enables single-flake electrochemical (de)lithiation of $MoS_2$ (a,b)** Microelectrode array chip for single-flake electrochemical cycling. $MoS_2$ flakes are exfoliated onto microelectrodes that act as the working electrodes (WE). A large $Li_{0.6}FePO_4$ porous electrode serves as the counter electrode (CE) and reference electrode. **(c)** Optical microscopy image of a $MoS_2$ flake as exfoliated on the working electrode (top) and after the ten electrochemical cycles (bottom) from **(d)**. $MoS_2$ flake was placed on an exposed Au microelectrode, while the surrounding area was passivated with a silicon nitride insulator. **(d)** Electrochemical cycling of a 2H-$MoS_2$ flake at a constant current of -3.6 and +2.4 pA. The first lithiation cycle features a voltage plateau around 1.1 V vs. $Li/Li^+$, while subsequent lithiation cycles do not show this plateau.

## II. RESULTS

†Contact author: yiyangli@umich.edu

## A. Single-flake Electrochemical Cycling of $Li_XMoS_2$

To perform single-flake electrochemical cycling on $MoS_2$, we utilize the microelectrode array developed in our previous works for single-particle cycling [42–44] (Figure 1a-b). $MoS_2$ flakes were exfoliated using Polydimethylsiloxane (PDMS) onto the 15-μm-square, 50-nm-thick Au microelectrodes (Figure 1c). Each microelectrode contains a single $MoS_2$ flake (~15 μm in size and ~80 nm thick) and serves as the working electrode in the electrochemical cell. The counter electrode slurry contains a larger $Li_{0.6}FePO_4$ porous electrode placed on large 2.5 mm square electrode at the center of the chip. $Li_{0.6}FePO_4$ has a stable voltage plateau ~3.4V vs. $Li/Li^+$ due to the phase transition and serves as a reliable reference and counter electrode [45]. The electrolyte is 1M $LiPF_6$ in anhydrous propylene carbonate and ionically connects the counter electrode and the working microelectrodes. More details can be found in the Experimental Methods section.

We performed electrochemical cycling on the exfoliated $MoS_2$ flake depicted in Figure 1c. We first pre-cycled the flakes by conducting 3 cyclic voltammetry cycles between 3.5 V and 1.3 V to prevent the parasitic reduction reactions with the liquid electrolyte from influencing future measurement (Figure S1, Supplemental Material); no phase transformation or cathodic currents is expected or observed at these voltages. Next, we conduct 10 electrochemical cycles on the flake over about 2 days by applying a constant current of -3.6 pA for lithiation and +2.4 pA for delithiation between the cutoff voltages of 0.8 V and 3.4 V (Figure 1d). During the first lithiation, an extended voltage plateau was observed at around 1.1 V. In subsequent cycles, this long plateau diminished. In contrast, the voltage profiles for delithiation remain relatively constant throughout the cycling, which closely monitors the lithiation profiles past the first cycle but in reverse. Our single-flake electrochemistry voltage profiles are largely consistent with previous electrochemical cycling of $MoS_2$ porous electrodes [31,36,46]. Using this platform, we achieve single-flake electrochemical (de)lithiation $MoS_2$ conducted at constant current for multiple cycles for the first time.

## B. Electrochemical and Structural Analysis of Phase Transitions in $Li_XMoS_2$

We next conduct a more detailed analysis of the single-flake electrochemistry. Figure 2a plots the overlayed voltage profiles of lithiation and delithiation over the same 10 cycles from Figure 1d. During the first cycle, when the lithium ion is inserted into the van der Waals gaps, it induces a phase transition from the 2H to the 1T phase, as signified by the long voltage plateau around 1.1 V (Figure 2a) and consistent with previous studies [30,35,47]. During lithium removal, no cathodic voltage plateau is observed near 1.1 V. Instead, delithiation progressed at much higher voltages, with cathodic peaks around 1.9 V and 2.4 V in the dQ/dV plots in Figure 2b. Because the open-circuit voltage represents the change in the Gibbs Free Energy upon the addition of lithium, this substantial difference in the voltage profile suggests that a lithiation-delithiation cycle is not fully reversible. In other words, the delithiated flake is different from the initial pristine flake. As we later show through Raman characterization, our results suggest that the flake has transformed into the 1T phase during lithium insertion and does not immediately revert to the 2H phase even after lithium is removed. In subsequent cycles, the 1.1 V voltage plateau largely diminishes and eventually disappears over time; instead, lithium insertion again occurs above 1.7 V. Because the peak at 1.1 V disappears, this result provides further evidence that lithium is not being inserted into the 2H phase in subsequent cycles like it was in the first cycle.

Figure 2c quantifies the change in the integrated lithiation capacity below and above 1.5 V over the ten cycles. On the first cycle, lithiation primarily occurs below 1.5 V, consistent with a 2H to 1T phase transition. In subsequent cycles, however, there is more lithiation capacity above 1.5 V. This result again shows that the 2H to 1T phase transition primarily occurs during lithiation on the first cycle, and that subsequent cycles do not induce a substantial 2H to 1T phase transformation.

Next, we conduct Raman spectroscopy on the flake before and after electrochemical cycling. The pristine flake shows clear $E_{2g}$ and $A_{1g}$ peaks at 384 and 410 $cm^{-1}$. However, after ten lithiation and delithiation cycles, we further observe a Raman peak at 156 $cm^{-1}$. This peak is consistent with previous reports of 1T-$MoS_2$ and not observed in 2H-$MoS_2$ [15,48–50]. Both the electrochemical voltage profile and Raman spectroscopy shows that we could synthesize a delithiated 1T-$MoS_2$ flake using electrochemical methods.

†Contact author: yiyangli@umich.edu

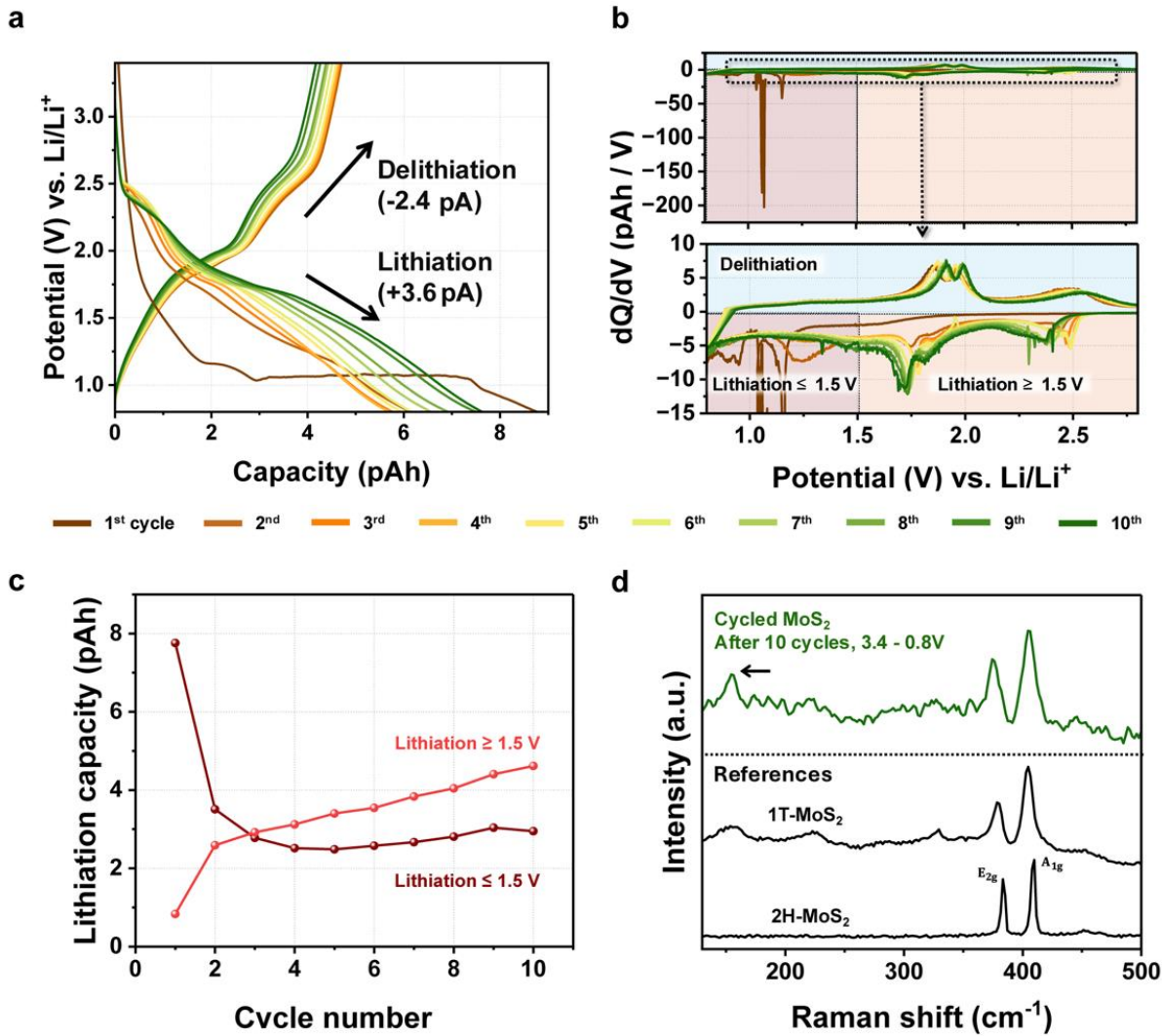

**Figure 2. Phase transitions upon electrochemical (de)lithiation in MoS₂** **(a)** Electrochemical lithiation and delithiation voltage profile for Li$_x$MoS$_2$ over ten cycles at constant current **(b)** The dQ/dV plot of this constant current cycling. The upper image presents the entire y-axis range, while the bottom image shows a more zoomed y-axis from -15 to 10 pAh/V. The potential profiles evolve with each cycle, indicating an irreversible phase transition from the 2H to the 1T phase. **(c)** The integrated lithiation capacities above and below 1.5 V show greater lithiation capacities occurs above 1.5 V over time. **(d)** Raman spectra of as-exfoliated 2H-MoS₂ before lithiation, 1T-MoS₂ synthesized by chemical lithiation with n-butyllithium solution and rinsed by deionized water, and our cycled and delithiated MoS₂. The Raman spectrum of 1T-MoS₂ was obtained from reference [48].

Summarizing the experimental results thus far, our electrochemical profiles show that pristine 2H-MoS₂ transforms into the 1T phase upon the insertion of lithium, as signified by a voltage plateau at 1.1 V. However, this voltage plateau is not present upon delithiation or subsequent lithiation cycles. Instead, lithium insertion in subsequent cycles occurs at higher voltages, with the low-voltage peak corresponding to the 2H to 1T phase transition rapidly diminishing. Correlative Raman spectroscopy suggests that the flake remains in the 1T phase and does not transform back into the 2H phase during subsequent lithium insertion and removal. We note that we are unable to distinguish between a regular or distorted 1T phase, as reported in some experiments [51,52].

### C. Delayed 1T to 2H Phase Transition in MoS₂

Having shown that the cycled MoS₂ flake remains in the 1T phase, we next assess the stability of this phase over time. To do so, we conduct Raman spectroscopy on both lithiated and delithiated MoS₂ flakes, both of which were cycled for two full cycles electrochemical at constant currents of -3.5 pA for lithiation and +2.5 pA for delithiation (Figure S2, Supplemental Material). Afterwards, one flake was again lithiated at -3.5pA to 0.8V (Figure 3a), and the current stopped while the flake was lithiated. A second flake was fully cycled a third time with a final potential of 3.4V (Figure 3b). After rinsing the liquid electrolyte with anhydrous dimethyl carbonate [53], the chip was stored in an Ar glovebox over several days. Raman spectroscopy was taken 30 minutes, 4 days, and 8 days after electrochemical cycling; the sample was stored in an Ar glovebox except during the Raman measurements.

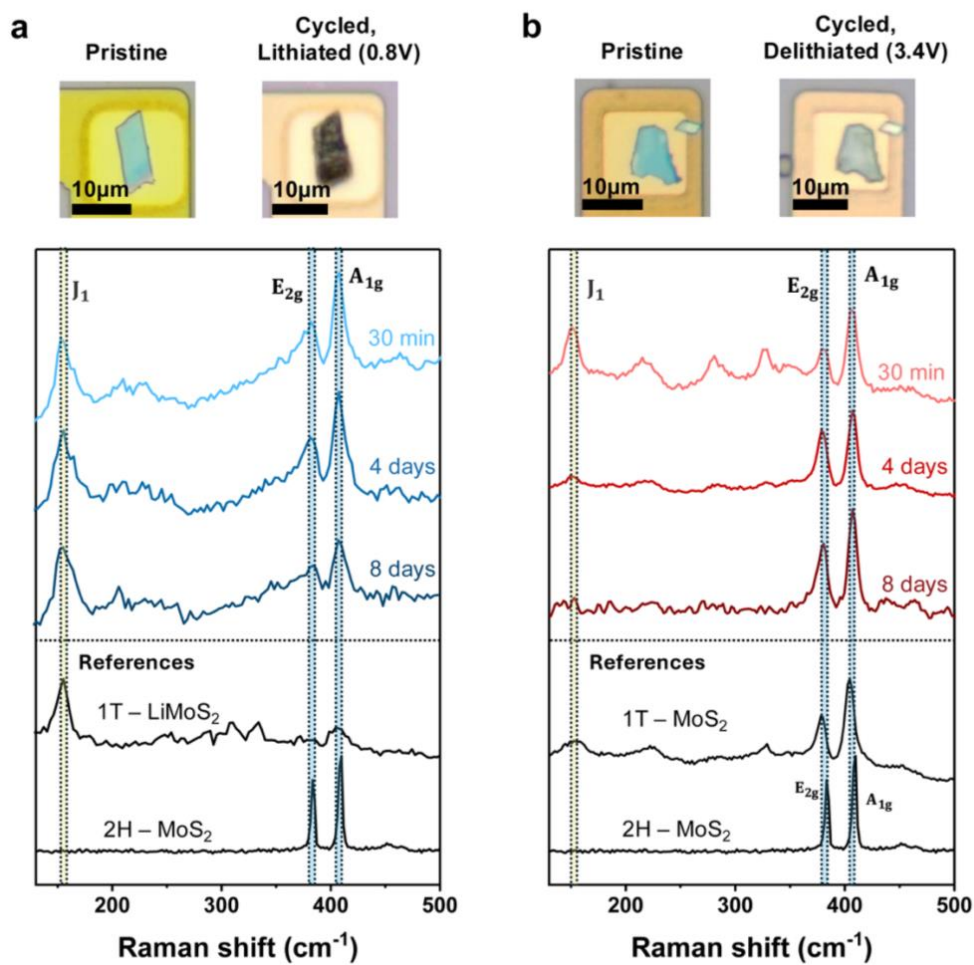

**Figure 3.** Evolution of the Raman spectra over time in electrochemically cycled $MoS_2$ flakes. Each spectrum was obtained 30 minutes, 4 days, and 8 days after cycling. **(a)** Lithiated $MoS_2$ flake cycled to 0.8V show that the flake remains in the 1T phase. **(b)** Delithiated $MoS_2$ flake cycled to 3.4V shows a progression back to the 2H phase. The reference if 1T-LiMoS2 was taken from a MoS2 flake chemically lithiated with n-Butyllithium for 1 day. Raman spectrum of 1T-$MoS_2$ in (b) was obtained from reference [48].

We first analyze the lithiated flake (Figure 3a). Raman spectroscopy shows that this flake has substantially transformed into the 1T phase upon electrochemical cycling, with strong $J_1$ peaks at 156 $cm^{-1}$, and red-shifted $E_{2g}$ and $A_{1g}$ peaks. Over the next eight days, this flake retains its 1T character. The $E_{2g}$ and $A_{1g}$ intensity is stronger than that of the chemically lithiated reference, but decreases in intensity over time; this result suggests that the lithiated flake has some remaining 2H character and becomes more 1T over time. Despite the likely presence of mixed phases, this result is consistent with the 1T phase being more stable for lithiated LiMoS2 [32,48,54].

Next, we consider the delithiated $MoS_2$ flake cycled to 3.4V (Figure 3b). This flake initially shows a strong $J_1$ peak at 156 $cm^{-1}$ after electrochemical cycling, similar to the flake in Figure 2d, and consistent with the 1T phase, although likely with some 2H remaining [48]. However, this $J_1$ peak weakens over eight days while the $E_{2g}$ and $A_{1g}$ become

†Contact author: yiyangli@umich.edu                                                                                      5

stronger and slightly blueshift back. This result shows that a delithiated flake slowly transforms back into the 2H phase over time. The 2H phase is the thermodynamically stable configuration [55]. This slow reversion from 1T back into 2H was also observed by Guo et al. after chemically lithiated $Li_XMoS2$ was subsequently delithiated by oxidizing in water [38]; our work shows a similar behavior in electrochemically cycled and delithiated $MoS_2$ in the absence of chemical oxidants.

Having shown that the electrochemically cycled and delithiated $MoS_2$ flake reverts from the 1T to the 2H phase over time with Raman spectroscopy (Figure 3b), we consider how this affects the electrochemical profile. Here, instead of constant current, we utilize cyclic voltammetry at a scan rate of 2 mV/s (Figure 4a); this data looks similar to the dQ/dV plot previously shown (Figure 2b). In the first lithiation (anodic) sweep, a cyclic voltammetry peak is observed at 0.9 V and is associated with the 2H to 1T phase transformation. The lower voltage (0.9 vs 1.1 V) is a result of the electrochemical overpotential associated with the larger current (up to -120 pA vs -3.6 pA in Figure 2b). Upon delithiation, we observe cathodic peaks around 1.7 V. Over the next five cycles, the phase transition peaks below 1.5 V gradually decrease while the peaks above 1.5 V increase, again similar to the constant current results. By the fifth cycle, there is essentially no anodic lithiation peak below 1.5 V in the cyclic voltammetry.

After completing five cycles, which end at a delithiated voltage of 3.4 V, we rested the $MoS_2$ flake in the Ar-filled glove box covered with the electrolyte for 8 days before restarting the cyclic voltammetry measurements. Here, we see a lithiation peak re-emerge below 1.5 V in the first cycle after the eight-day rest (Figure 4b). This anodic peak has somewhat higher voltage with its height diminished compared to the initial sharp peak in the first cycle, possibly attributed to the expansion of van der Waals layers and the formation of wrinkles which may facilitate subsequent lithiation. The capacities below and above 1.5 V were calculated by integrating the lithiation current in the corresponding range (Figure 4c). The lithiation capacity below 1.5 V exhibited the highest value (9 pAh) at the first cycle and significantly dropped to below 3 pAh after the fifth cycle. However, after the resting period of 8 days, the lithiation capacity below 1.5 V recovered to 5 pAh in the first cycle after resting (6th cycle overall), which is a substantial increase from the 5th cycle (3 pAh) before the eight-day rest. We note that the lithiation capacity below 1.5 V cannot be 0 even in the absence of the 2H phase due to capacitive currents and solid solution intercalation.

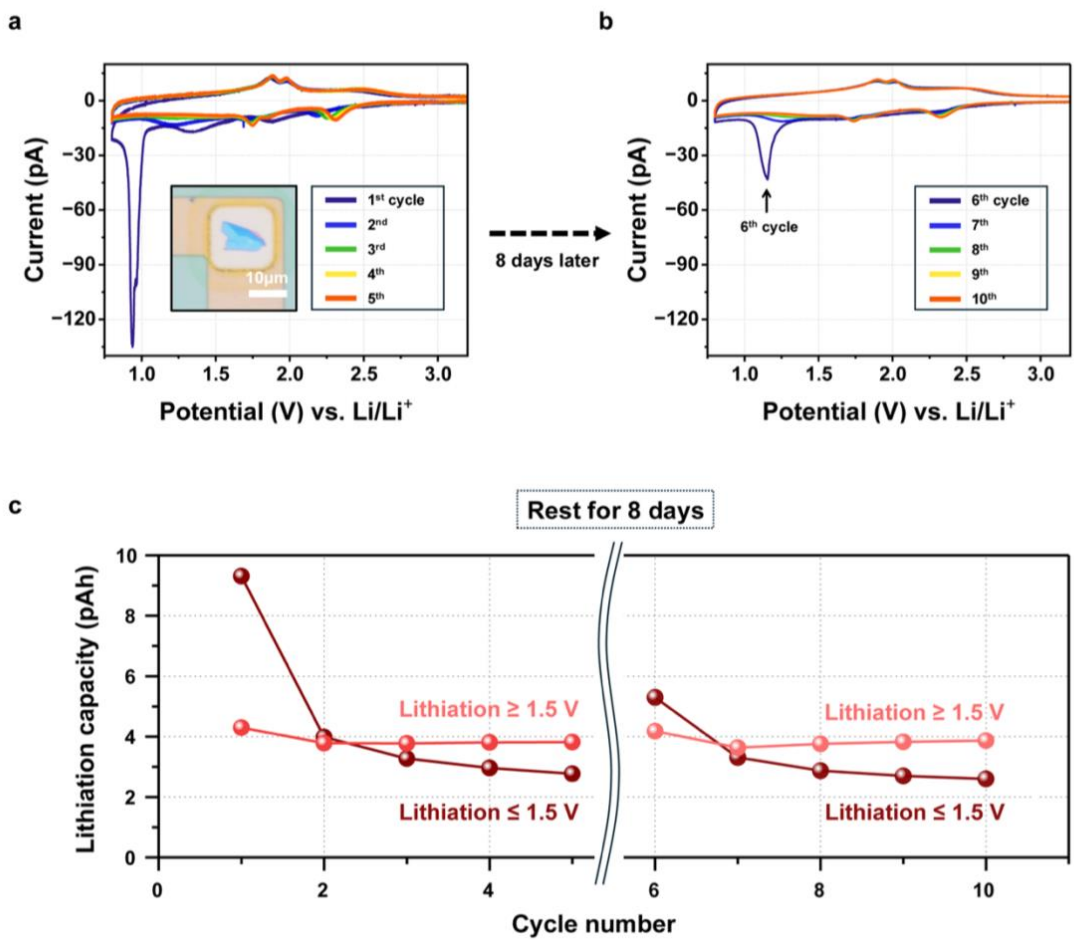

**Figure 4. Evidence of delayed phase transition from electrochemical voltage analysis. (a)** Cyclic voltammetry curves of the first set of lithiation and delithiation cycles of $MoS_2$ after exfoliation. **(b)** The second set of five cycles taken eight days after the first set of cycles ended. During these eight days, the cycled $MoS_2$ flake was covered with the electrolyte inside an Ar glovebox. **(c)** Integrated lithiation capacity shows a substantial increase in the <1.5 V lithiation capacity, consistent with the peak at 1.1 V in **(b)**.

†Contact author: yiyangli@umich.edu

Our combined Raman and cyclic voltammetry results both clearly show that the electrochemically delithiated 1T-MoS$_2$ flake would transform back into 2H over the course of several days. This result is consistent with the lower energy of the 2H phase compared to the 1T phase in MoS$_2$ [56,57]. However, it also shows that the 1T phase remains for several days due to the slow kinetics of the 1T to 2H phase transformation.

## III. DISCUSSION

Figure 5 shows a schematic summary of our findings. During electrochemical lithiation, 2H-MoS$_2$ transforms into the 1T Li$_X$MoS$_2$ phase. However, upon electrochemical delithiation, the MoS$_2$ flake does not instantly revert back into the 2H phase even when lithium has been removed. Instead, the 1T-MoS$_2$ flake is stable for several days before slowly reverting back to the 2H phase. While the flake is probably mixed phase throughout the relaxation process, our results show the general trend and direction between 2H and 1T phases during electrochemical cycling and during relaxation. A similar slow transformation from 1T to 2H was observed when a chemically lithiated Li$_X$MoS$_2$ flake was delithiated using water [38,39]. However, our work shows that this slow transformation could be achieved in electrolyte or inert argon. In contrast, water or other oxidants used to chemically remove lithium may induce other chemical reactions that affect the phase behavior.

Our results showing 1T-MoS$_2$ upon delithiation are most consistent with that of Cook, Tolbert, and colleagues [36,58,59], which also showed the 1T phase in electrochemically delithiated MoS$_2$. However, we not only conducted our measurements on single flakes, but also experimentally showed that the 1T phase reverts back to the 2H phase. Our results are different from single-flake results which show the 2H phase after electrochemical lithiation and delithiation [15,34]. One possible explanation is that our samples are much better sealed against possible leakage—we not only constructed a cell using silicon, glass, stainless steel, and fluoropolymer O-rings, but cycled this cell within an Ar glovebox. In contrast, previous works often conduct electrochemistry within heat-sealed plastic or epoxy-sealed housing in air , where greater water vapor and oxygen gas leakage is expected [8,15]. Side reactions caused by water and oxygen, such as surface oxidation [60], electrolyte decomposition [61] and formation of byproducts such as Li$_2$MoO$_4$ [54]. They could affect phase transformation dynamics in MoS$_2$, by directly oxidizing the flake or by accelerating the transformation back to 2H [62,63]. Finally, we limit our voltage to above 0.8 V vs Li/Li$^+$, before the conversion of MoS$_2$ into Mo metal and Li-S, which would maintain the microstructure of the flake during electrochemical (de)lithiation.

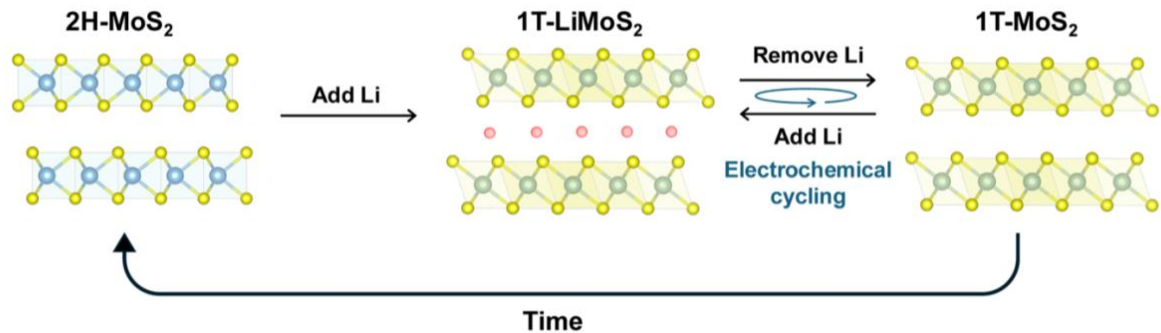

**Figure 5.** Schematic illustrations of the polymorphic phase transition in MoS$_2$. The initial lithiation and following repeated electrochemical cycling of lithiation and delithiation induces the phase transition from the 2H phase to the 1T phase.

†Contact author: yiyangli@umich.edu

By showing that $MoS_2$ remains in the 1T phase even after delithiation, our results have implications in the design of $MoS_2$-based electrochemical devices. For batteries, we show that the $MoS_2$ remains in the metallic 1T phase even after delithiation, which enables faster cycling rates if electronic conductivity is rate limiting [36,58]. In lithium insertion switching devices for neuromorphic computing [7,20,21,64] and thermal transistors [3,19], we anticipate that, beyond the first cycle, $MoS_2$ would switch dynamically between the delithiated and lithiated 1T state, not between the 2H and 1T phases. On the other hand, the slow transition from 1T to 2H may enable the stabilization of mixed phase 2D materials. Such mixed phase may contain interesting electronic states, such as endotaxial charge density waves recently observed in $TaS_2$ [65,66].

## IV. CONCLUSIONS

In summary, we investigated the structural phase transition in $MoS_2$ during electrochemical lithiation and delithiation. Through correlative electrochemical voltage analysis and Raman spectroscopy, we observed that the $MoS_2$ flake does not immediately revert back to the 2H phase upon lithium removal. Instead, this process takes several days. This result has implications for electrochemical $MoS_2$ devices as well as phase engineering of $MoS_2$ flakes.

## V. METHODS

**Microelectrode array fabrication**: All microelectrode array chips were fabricated at the Lurie Nanofabrication Facility (LNF) at the University of Michigan based on previous work [42–44]. The microelectrode array was fabricated on 100 mm silicon wafer with a 500 nm thermal oxide layer. To pattern the Au electrodes, we spin-coated 0.9 µm of LOR 10B and 1 µm of S1813 photoresist, exposed the photoresist to ultraviolet light from an aligner over a photomask, and developed with a Tetramethylammonium hydroxide developer. Next, we deposit a 5 nm / 100 nm of Ti/Au metal layer using electron beam evaporator on the patterned photoresists and immersed in the etchant (Remover-PG) to lift off the remaining photoresist and metal layer on top of it. Next, we deposit and pattern a passivation layer to minimize parasitic electrochemical reactions between the Au electrode and the electrolyte. A 50 nm of SiN layer is deposited on the patterned Ti/Au layer using plasma-enhanced chemical vapor deposition (Plasmatherm 790). The second photolithography process was conducted by applying 3 µm of SPR220 photoresist and development in tetramethylammonium hydroxide developer (mask pattern in Figure S3, Supplemental Material). Next, we used reactive ion etching (RIE) to etch the SiN passivation layer to open the Au in select locations to transfer the $MoS_2$ flakes. After etching, the remaining SPR220 was removed using a Plasma stripper, then the processed wafer diced into 3´3 cm chips using a dicing saw.

**$MoS_2$ flake exfoliation and assembly**: All $MoS_2$ flakes were exfoliated by using mechanical exfoliation method. A $MoS_2$ crystal purchased from HQ graphene was put on blue scotch tape and folded and unfolded multiple times to cleave the crystal to few-layer flakes. The $MoS_2$ flakes on the blue scotch tape were then transferred onto a PDMS polymer film. The array chip was placed on a table under a 20X optical microscope objective and an xyz manipulator. The PDMS film with $MoS_2$ flakes on top of it was placed on the coverslip and secured to the xyz manipulator to place the flakes onto the exposed Au regions on the array chip. After positioning the exfoliated $MoS_2$ flakes, the microelectrode array with flakes was annealed at 300℃ in air for 1 hour.

**Counter/Reference Electrode**: A slurry of partially delithiated $Li_{0.6}FePO_4$ mixed with PVDF and carbon black was used as the counter/reference electrode. $Li_{0.6}FePO_4$ was synthesized by partially delithiating a pristine carbon-coated commercial $LiFePO_4$ (Mitsui Engineering and Shipbuilding) in aqueous $K_2S_2O_8$ with a molar ratio of 1 $K_2S_2O_8$: 5 $LiFePO_4$. After chemical delithiation for about 1 day, the solution was centrifuged at 3000 rpm to separate the insoluble $Li_{0.6}FePO_4$ from the water-soluble $K_2S_2O_4$ and $Li_2S_2O_4$.

The partially delithiated $Li_{0.6}FePO_4$ was dried in an 80℃ vacuum oven overnight to completely remove the residual water. The $Li_{0.6}FePO_4$ was then mixed with polyvinylidene fluoride (PVDF) and carbon black with a 7:1:2 weight ratio, with N-methyl-2-pyrrolidone was used as a solvent. This slurry was drop-casted at the center of the microelectrode array and heated at 80℃ on hot plate to evaporate the N-methyl-2-pyrrolidone solvent.

**Electrochemical Cycling**: The microelectrode array was placed in a glove box filled with Ar gas for electrochemical cycling. 3 µl of 1M $LiPF_6$ in propylene carbonate was dropped at the center of the array to cover the $MoS_2$ flakes and a counter/reference electrode. The sealed electrochemical cell was sealed using a stainless-steel module with fluoropolymer O-rings and operated in an Ar-filled glovebox with <1 ppm O2 and H2O (Figure S4, Supplemental Materials). A Bio-Logic VMP300 potentiostat with ultra-low-current modules was used to perform all the electrochemistry. Tungsten needles on an XYZ manipulator were placed on the Au pads at the edge of the array to make an electrical connection to the individual flake.

†Contact author: yiyangli@umich.edu

The electrochemistry was conducted using constant current and cyclic voltammetry to insert and extract lithium into and out of the $MoS_2$ flake. In constant current measurements, we first conducted 3 or 3 cyclic voltammetry precycling at the scanning rate of 1 mV/sec with cutoff voltage at 1.1 V ~ 1.3 V to minimize parasitic side reactions from the first cycle upon electrolyte reduction at ~2.0 V (Figure S1, Supplemental Material). Afterwards, we conduct constant current measurements between 0.8 and 3.4V at a constant current at 3.5 pA for lithiation and 2.5 pA for delithiation. For the cyclic voltammetry, the voltage was swept from 3.4 V to 0.8 V at a scan rate of 2 mV/s after precycling once between 1.8 V and 3.4 V at 2 mV/s.

**Raman Spectroscopy**: Raman spectroscopy was conducted using a Renishaw InVia micro-Raman spectrometer with a 532 nm laser source under 50× long working distance objective. After electrochemical cycling, the microarray chips were rinsed with anhydrous dimethyl carbonate and subsequently analyzed by Raman spectroscopy. The relative Raman power was 0.5%.

## ACKNOWLEDGMENTS

This research was carried out under the auspices of the Center for Materials Innovation, a Materials Research Science and Engineering Center at the University of Michigan, supported by the National Science Foundation, Award No. DMR-2309029. The microelectrode arrays were fabricated at the University of Michigan Lurie Nanofabrication Facility.